\documentstyle[aps,multicol]{revtex}
\begin{document}

\title{Fractal formation and ordering in random sequential adsorption}
\author{N.~V.~Brilliantov$^{1,2}$, Yu.~A.~Andrienko$^{1,3}$,
P.~L.~Krapivsky$^{4}$, and J.~Kurths$^{3}$}
\address{$^{(1)}$Moscow State University, Physics Department,
Moscow 119899, Russia}
\address{$^{(2)}$Department of Chemistry, University
of Toronto, Toronto, Canada M5S 1A1}
\address{$^{(3)}$University Potsdam, Physics Department, Am Neuen Palais, 
D--14415 Potsdam, Germany}
\address{$^{(4)}$Courant Institute of Mathematical Sciences,
New York University, New York, NY 10012-1185, USA}
\maketitle

\bigskip
\begin{abstract}

We reveal the fractal nature of patterns arising in
random sequential adsorption of particles with continuum power-law
size distribution, $P(R)\sim R^{\alpha-1}$, $R \le R_{\rm max}$.  We
find that the patterns become more and more ordered as $\alpha$ increases,
and that the Apollonian packing is obtained at 
$\alpha \to \infty $  limit. We introduce the entropy production rate as a
quantitative criteria of regularity and observe a transition from
an irregular regime of the pattern formation to a regular one. We
develop a scaling theory that relates kinetic and 
structural properties of the system. 

\end{abstract}

\begin{multicols}{2}

A variety of physical, chemical, and biological problems can be
modeled by random sequential addition (RSA) processes
\cite{privman,evans}. Examples include adhesion of proteins and
colloidal particles onto surfaces, polymer chain reactions, car
parking, etc.  The structure of disordered media, e.g.
non-crystalline and granular materials, is also studied within the
RSA-type model approaches (cf. \cite{medved,pavlov}).  In the simplest
form, RSA processes can be formulated as sequential addition of
objects that cannot overlap and, once inserted, cannot move or leave
the structure.

Geometric and kinetic characteristics of RSA processes with identical
objects are fairly well-known \cite{privman,evans}.  In contrast,
adsorption of mixtures has been addressed in a very few studies
\cite{baker,talbot,bartelt,meakin,bonnier,pk}.  If a mixture contains
a small number of different sizes, geometric and kinetic
characteristics are primarily determined by the smallest size. In some
applications, e.g. for adsorption of latex spheres, or in modeling the
structures of the coal grinds, ceramic or metallic powders, the size
distribution may spread for several decades \cite{medved,coal}.
Therefore, before the smallest size will finally win, an interesting
intermediate asymptotics arises. To address this intermediate regime,
we consider a {\it continuous} distribution with sizes ranging from
zero up to some maximal size which we set equal to unity. 
For definiteness, we choose a power-law form

\begin{eqnarray}
\label{distrib}
P \left( R  \right)  =
\left\{\vcenter{
\hbox{ \hbox to 15mm  {$\alpha \, R^{\alpha -1}$ \hss} for $ R \le 1$,}
\hbox{ \hbox to 15mm  {$0$\hss} for $R >1$,}}
\right.
\end{eqnarray}
for the rate of adsorption of particles with radii in the interval
$(R,R+dR)$ ($\alpha > 0$ due to the normalization requirement).  In
the present Letter we show that RSA of particles with the size
distribution (\ref{distrib}) gives rise to fractal structures with a
fractal dimension that depends on the exponent $\alpha$. We also show
that in the large $\alpha$ limit, highly regular structures arise
which tend to the famous Apollonian packing as $\alpha \to \infty$.

Let $\Phi(t)$ is the fraction of uncovered area at time $t$ and
$\Psi(R,t)$ is the probability that a disc of radius $R$ can be placed
onto the surface, with a center at some arbitrary point.  The
definition of $\Phi(t)$ and $\Psi(R,t)$ implies $\Psi(0,t)=\Phi(t)$
and
\begin{equation}
\frac{d\Phi }{dt}=-\int_0^\infty dR P(R)\Psi (R,t)\Omega_dR^d.
\label{phi1}
\end{equation}
Eq.~(\ref{phi1}) is written for the general case of adsorption
onto a $d$-dimensional substrate, and $\Omega_d$ denotes the volume of
the $d$-dimensional unit ball.

For the size distribution
(\ref{distrib}) one can assume a scaling behavior for the function
$\Psi(R,t)$, i.e.
\begin{equation}
\Psi (R,t)=S^{\theta }F\left(\frac R{S(t)}\right).
\label{psi2}
\end{equation}
Here $S(t)\sim t^{-\nu }$ is a typical gap between neighboring
adsorbed particles, and the scaling description should be relevant in
the scaling regime, $t\gg 1$ and $R\ll 1$ with $R/S(t)$ finite.
Without loss of generality we can set $F(0)=1$ and thus we get $\Phi
(t)=S^{\theta}(t)$.  Therefore, the scaling assumption for $\Psi(R,t)$
implies $\Phi (t)\sim t^{-z}$ with $z=\theta\nu$. Expressing
Eq.~(\ref{phi1}) in the scaling form yields $\nu=(\alpha+d)^{-1}$ and
$z \simeq \alpha\Omega_d~ \int_0^\infty dx\, x^{\alpha+d-1} F(x)$.
Hence for $d=1$ we have $\nu=(1+\alpha)^{-1}$ in agreement with the
exact result of Ref.~\cite{pk}, where it was justified that for the
power-law distribution (\ref{distrib}), the scaling behavior does
appear in 1D.

Now we relate the geometric (fractal) properties of the arising
patterns with the kinetics.  Let $n(R)$ be a number of adsorbed
particles per unit area, with radii ranging from $R$ to $R+dR$.
Clearly,
\begin{equation}
n(R)=\int_0^\infty dt\,P(R)\Psi (R,t) \sim R^{\alpha-1+(z-1)/\nu }
\label{nr}
\end{equation}
We can determine the fractal dimension $D_f$ of the pore space by
introducing a cutoff size $\epsilon$, and calculating the number of
particles per unit volume, $N(\epsilon)$, with radii greater than
$\epsilon$. When this number behaves as a power law at small sizes
limit, i.e.,
\begin{equation}
N(\epsilon )=\int_\epsilon ^\infty dR\,n(R) \sim \epsilon ^{-D_f}
\label{neps}
\end{equation}
one concludes that the fractal dimension is $D_f$ (see
e.g. ~\cite{manna}).  Notice that the porosity,
i.e. the fraction of uncovered area, behaves as
$\epsilon ^{d-D_f}$.
Combining Eqs.~(\ref{nr}) and (\ref{neps}),
we arrive to the following result for the fractal dimension:
\begin{equation}
D_f=d-z(d+\alpha)
\label{df}
\end{equation}
Eq.~(\ref{df}) shows that the fractal dimension of
arising patterns is intimately related to the exponent $z$, which
describes the pattern formation kinetics. Similar relations between
kinetics exponents and spatial properties of the patterns
were observed for other pattern formation models ~\cite{and94,bril94}.
The exact result for $z$ in 1D ~\cite{pk} completely solves the 
problem for the one-dimensional systems.

For $d\geq 2$ we could neither find the exact value of the exponent
$z$, nor strictly prove the existence of scaling. To check the
validity of the scaling theory, we performed numerical study of the 2D
adsorption process.  For every value of $\alpha$ we generated $10^6$
discs, and then computed the fractal dimension from the relation
$N(\epsilon)\sim \epsilon^{-D_f}$ ~\cite{manna,and94,bril94}.  Typical
patterns for small and large values of $\alpha$ are shown on Fig.~1
and Fig.~2, respectively.  Fig.~3 gives the fractal dimension as a
function of the exponent $\alpha$.  We also present the value of
fractal dimension, calculated from the scaling relation of
Eq.~(\ref{df}), where numerical results for the kinetic exponent $z$
were used.  Fig.~3 indicates that the scaling theory works fairly well
in the whole range of $\alpha$.

The lack of evident spatial correlations at small $\alpha$ (see
Fig.~1) suggests that a mean-field theory (MFT) can provide a
reasonable approximation in this region.  We develop a MFT similar to
the one of Refs.\cite{and94,bril94} and derived the following equation
for the scaling function $F(x)$:

\begin{eqnarray}
\label{F1}
&&F(Rt^{\nu})= \\
&&\exp \left\{ -\int_0^td\tau \int_0^\infty d\rho P(\rho )\Omega_d
\left((R+\rho )^d-\rho ^d\right) F(\rho\tau^{\nu})\right\}. \nonumber
\end{eqnarray}

The ansatz $F(x)=\exp(-A_1x-\ldots -A_dx^d)$ solves Eq.~(\ref{F1}) in
arbitrary dimension.  In particular, in 2D the coefficients $A_1$ and
$A_2$ are determined from equations

\begin{eqnarray}
\label{aa}
A_1=2\pi\left( \alpha +2 \right) \alpha\, \int_0^{\infty} x^{\alpha}e^{-A_
1x-A_2x^2}dx \nonumber \\
A_2=\frac{\pi}{2} \left( \alpha +2  \right) \alpha\, \int_0^{\infty}
x^{\alpha-1}e^{-A_1x-A_2x^2}dx
\end{eqnarray}
Solving Eqs.~(\ref{aa}) and then inserting $F(x)=e^{-A_1x-A_2x^2}$
into expression $z=\alpha\pi\int_0^\infty dx x^{\alpha+1}F(x)$,
derived previously, one finds $z$ and $D_f$.  In particular,
$D_f=2-\pi\alpha-(\pi^{3/2}+\pi/2)\alpha^2+\ldots$ in the small
$\alpha$ limit.  Mean-field results are given in Fig.~3.  Fig.~3
indicates that the MFT works quite well for small $\alpha$, and it can
be shown that MFT becomes exact at $\alpha \rightarrow 0$ limit. It is
failed, however, for large $\alpha$, where the spatial correlations
seem to be very important and the arising patterns strongly resemble
the regular structures (see Fig.~2).

To quantify the increasing regularity of the structures, we introduce
an entropy, $S_N$, characterizing the degree of order of $N$-particles
patterns. If $C_k(N)$ denotes such a pattern and $p(C_k)$ denotes the
probability of that pattern, one can define the Shannon entropy
\cite{mackey,bla}:

\begin{equation}
S_N = -\sum_{C_k} p(C_k) \log_2 p(C_k)
\label{entdef}
\end{equation}
As it follows from the Eq.~(\ref{entdef}), $S_N=0$ for a regular
pattern, since only one definite configuration with the probability
$p=1$ contributes to the entropy.  $S_N$ rapidly increases with
increasing number of possible configurations, i.e.  with decreasing
order.  The closely related value, $ dS_N/dN \simeq S_{N+1}-S_N$ gives
the entropy production rate and characterizes the regularity of the
pattern formation process.

To find the entropy production rate, we first compute the conditional
entropy $S_N[C_k(N-1)]$ of the $N$ discs for the given pattern
$C_k(N-1)$ of $N-1$ discs. The conditional entropy is determined and
calculated as follows: The $N$th disc added to the pattern $C_k(N-1)$
can be treated as a point in the configuration $(x,y,R)$-space, where
$x$ and $y$ are coordinates of the disc center, $R$ the disc radius.
We divided the continuous configuration space into sufficiently small
discrete cells and enumerated these cells (for computations, we used
$\sim 10^{7}$ cells). Then, we introduced the probabilities $p_i$ that
the next disc comes to the $i$th cell for the given configuration
$C_k(N-1)$, and calculated the conditional entropy as
$S_N[C_k(N-1)]=-\sum_ip_i\log_2 p_i$.  The probabilities $p_i$ can be
determined numerically for any given configuration $C_k(N-1)$ of $N-1$
discs.  Namely, $p_i=A_iR_i^{\alpha-1} /\sum_jA_jR_j^{\alpha-1}$,
where $A_i$=0 if the disc corresponding to the $i$th cell overlaps
with some disk in the pattern $C_k(N-1)$, otherwise $A_i=1$.

The definitions of the conditional and full entropy then allows us to
find the entropy production rate, $S_{N+1}-S_N$, by performing the
averaging of the conditional entropy $S_N[C_k(N-1)]$ over all possible
configurations $C_k(N-1)$ of $N-1$ disks. In practice, the averaging was
performed over a number of ($\sim 10^{2}$) Monte-Carlo runs. 

To compare the entropy production rate for different values of
$\alpha$, we plot $dS_N/dN$ as a function of the free area $\Phi$ (see
Fig.~4).  Fig.~4 shows the striking behavior
of the entropy production rate at very large $\alpha$: At the
beginning of the process of pattern formation (i.e. at $\Phi \approx
1$), it decreases slowly in the same manner as for small $\alpha$, but
at $\Phi \approx 0.55$ a sharp decay to a (small) plateau value is
observed. One can interpret such type of behavior as a transition from
a regime of "low regularity" to a regime of "high regularity" in the
pattern formation process.  The threshold value, $\Phi \approx 0.55$,
is close to the jamming density, $\Phi_{\infty}=0.542\ldots$, of the
ordinary RSA of identical discs on a plane.  The transition can be
understood as follows: At the first "non-regular" stage of the
process, random patterns mainly composed of discs with $R=1$
arise. After the jamming limit is achieved, the rule "insert the disc
of maximal possible radius" starts to work [as it follows from
Eq.~(\ref{distrib}) in the limit $\alpha \to \infty$], and much more
regular patterns are produced.  This transition corresponds to the
sharp decrease of the entropy production rate observed in Fig.~4.

Now we argue that the fractal dimension of the patterns
arising at the limit $\alpha \rightarrow \infty $ tends to that of the
famous Apollonian packing, the oldest known fractal construction 
(see \cite{man}). To realize this we notice that 
in accordance with the rule "insert the disc of the maximal 
size", that works at $\Phi < \Phi_{\infty}$, 
the every new added disc touches (at least) three other ones.
Therefore, after some time, all the free area will be scattered 
into a finite number (say $L$) of curvilinear triangles, confined by 
discs of radii $(a_l, b_l, c_l)$, $l=1, \ldots, L$. Contribution 
to the free area from the other curvilinear polygons (confined by 
4, 5, etc., touching discs) is negligible if time is large enough. 
Filling of these uncovered curvilinear triangles is performed in the 
very same way as in the Apollonian packing, namely, the every next 
disc placed touches from the inside the three outer touching discs. 
The fractal dimension of the pore space, obtained by the Apollonian
filling does not depend on the starting set $(a_l, b_l, c_l)$ and
equals to $D_A=1.305\ldots$  ~\cite{man}.
Thus we conclude that the fractal dimension of the patterns arising at
$\alpha \rightarrow \infty $ coincides with that of the Apollonian
packing. 

In summary, we have investigated the adsorption kinetics and spatial
properties of the arising patterns in a random sequential adsorption
of particles with power-law size distribution. We have developed a
scaling approach and verified scaling description by comparing with
exact results in one dimension and numerical results in two
dimensions.  We have found that arising patterns have the fractal
dimension $D_f$ that sensitively depends on the power-law exponent of
the particles size distribution and is intimately related to the
kinetic exponent of the time-dependent coverage. We have deduced that
when $\alpha$, increases from $0$ to $\infty$, $D_f$ decreases from 2
to $D_A=1.305\ldots$, of the Apollonian packing. 
We have observed that the regularity of patterns increases with
increasing $\alpha$ and introduce the entropy production rate as a
quantitative criteria of the regularity.


\bigskip\bigskip

Fig.~1  A typical pattern for $\alpha=0.1$.  Only a small part of the 
        total number of discs is shown.

Fig.~2  The same as in Fig.~1,  $\alpha=50$.

Fig.~3  The fractal dimension $D_f$ versus $\alpha$. In the inset:
        kinetic exponent $z$ versus $\alpha$. The mean-field results 
        are obtained from numerical soluion of Eqs.~(\ref{aa}).

Fig.~4  The entropy production rate versus the uncovered area $\Phi$ 
        for different values of $\alpha$.

\end{multicols}

\begin{references}
\bibitem{bla}
        The summation over a continuous set of configurations implicitly
        assumes that we consider a finite subset of configurations; 
        in a numerical implementation, this is clearly the case.
\end{references}
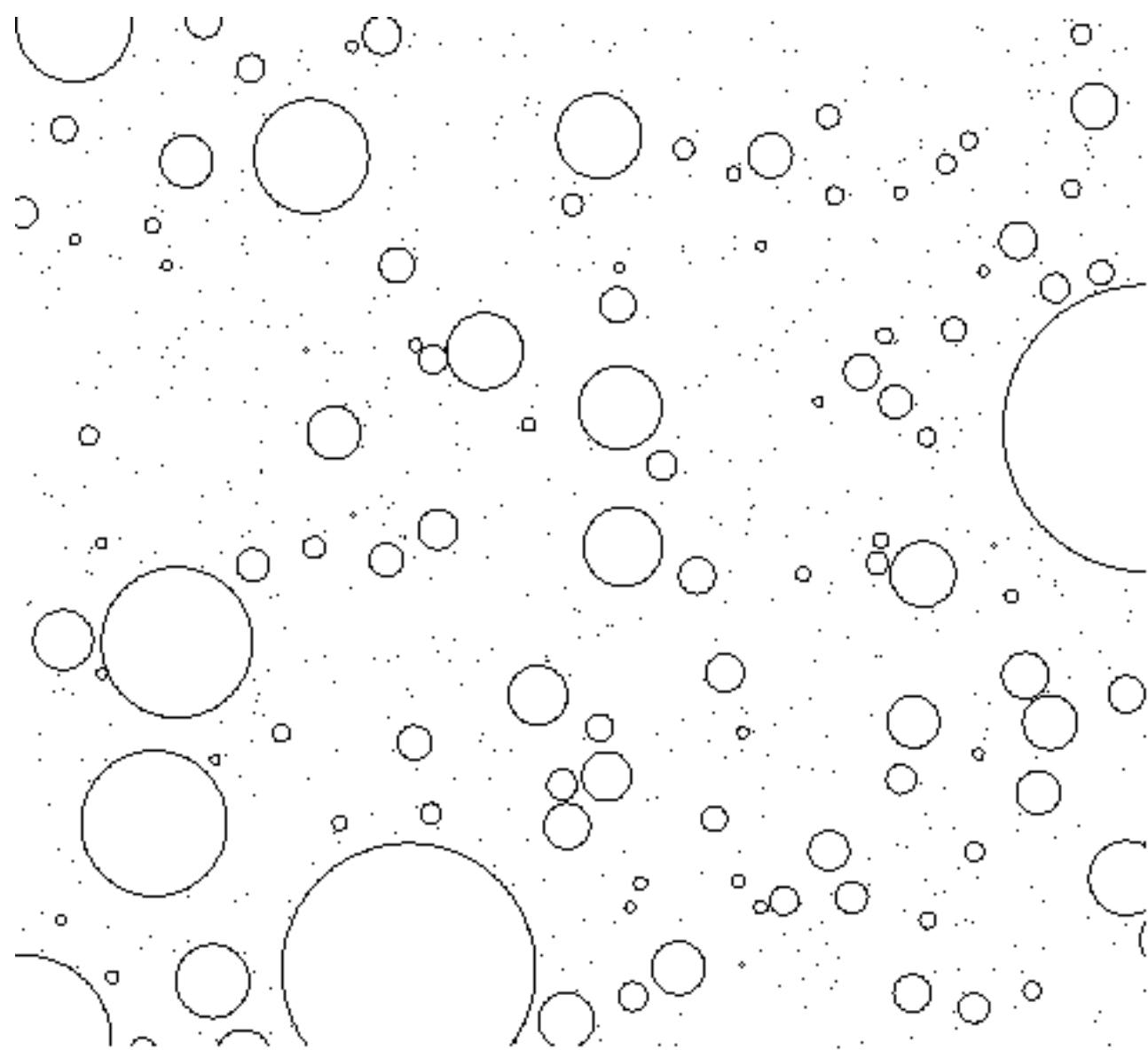
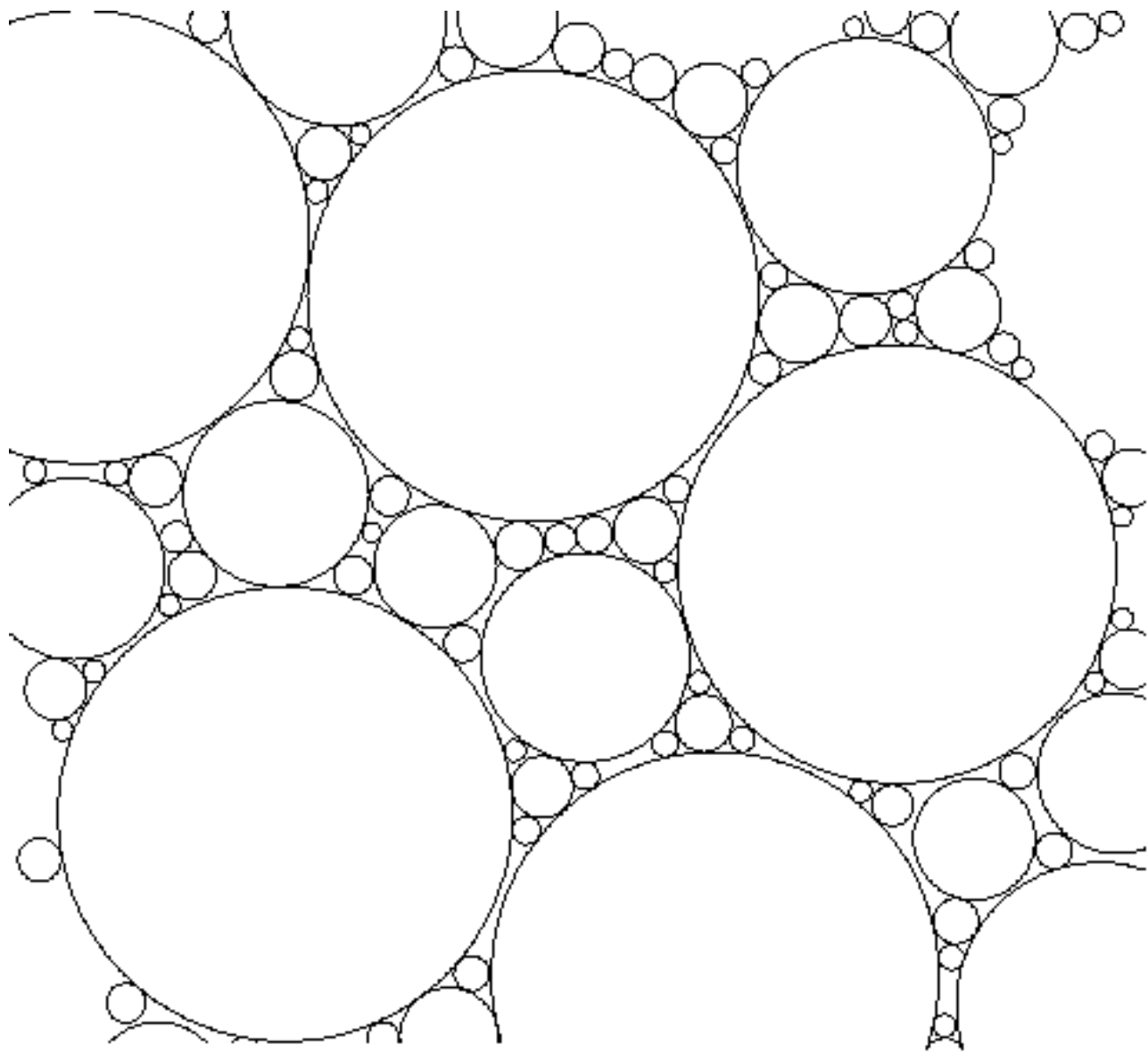
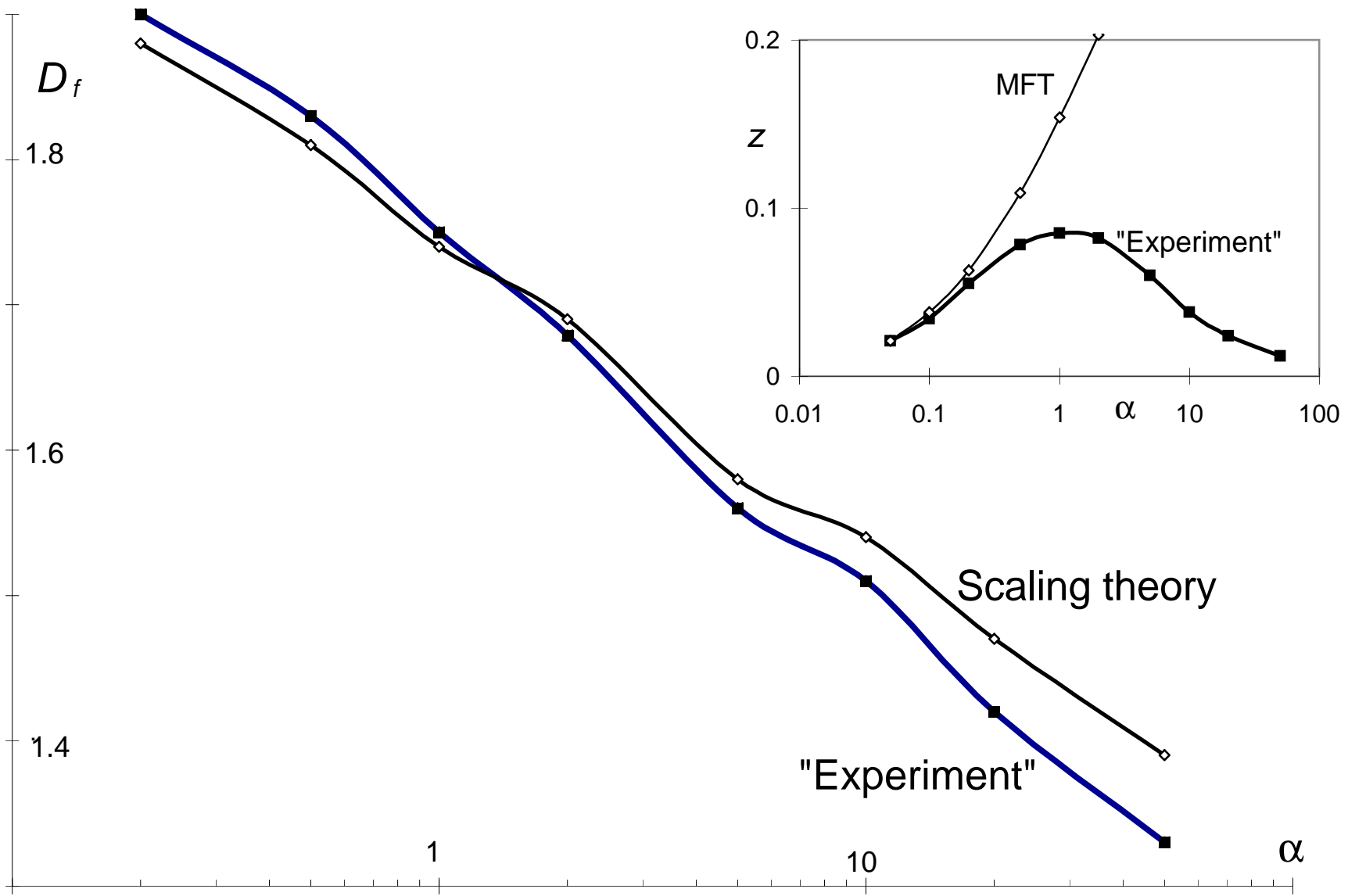
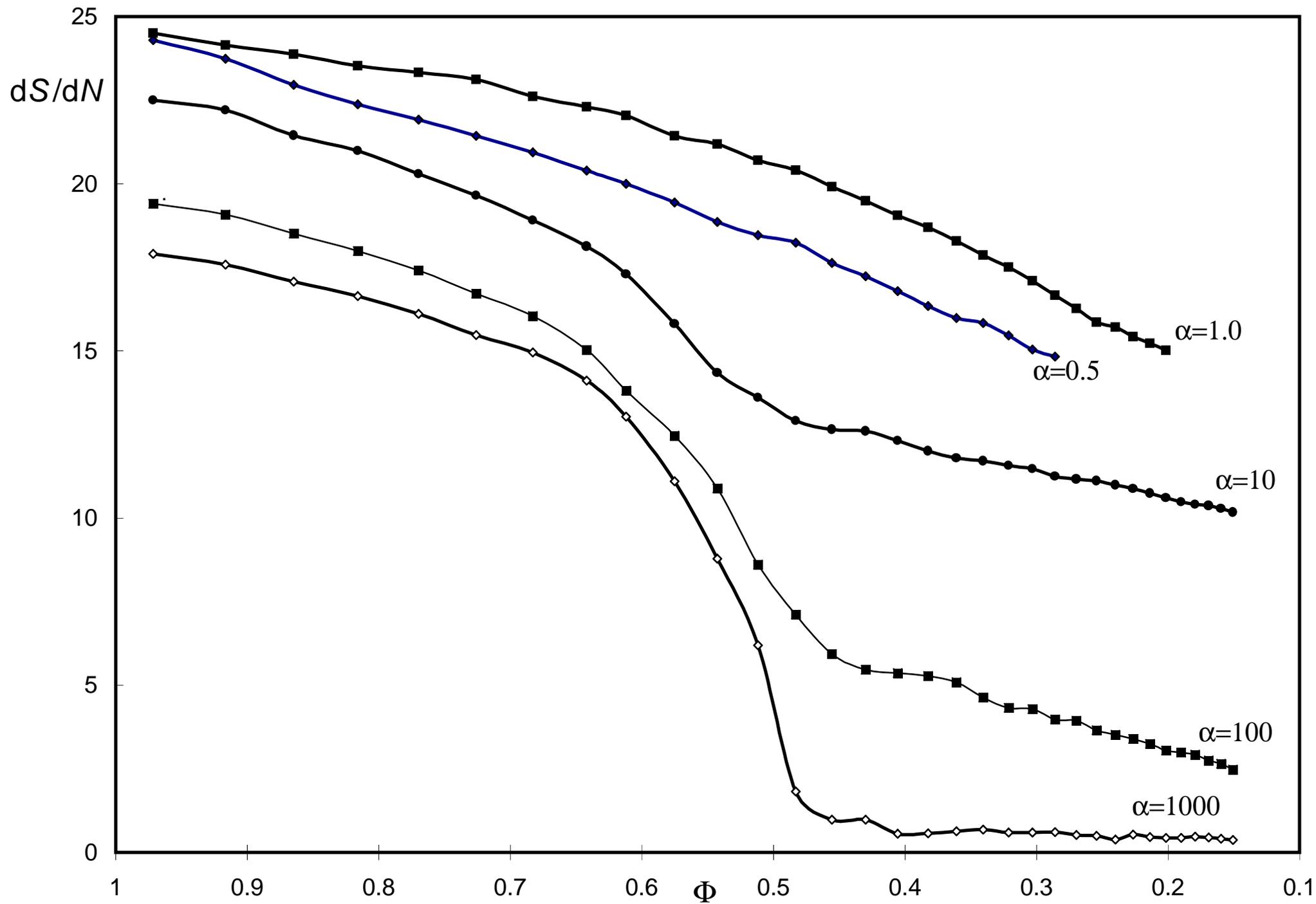
\end{document}